\documentclass[aps,pra,twocolumn,groupedaddress]{revtex4-2}
\usepackage{amsmath, amsthm, amscd, amsfonts, latexsym, amssymb, graphicx, color}
\usepackage{graphics}
\newcommand{\bs}{\boldsymbol}
\newcommand{\beqa}{\begin{eqnarray*}}
\newcommand{\eeqa}{\end{eqnarray*}}
\newcommand{\beqn}{\begin{eqnarray}}
\newcommand{\eeqn}{\end{eqnarray}}

\newcommand{\lt}{\left}
\newcommand{\rt}{\right}

\newcommand{\A}{\mathbb A}

\newcommand{\C}{\mathbb C}

\newcommand{\R}{\mathbb R}

\newcommand{\tf}{\tfrac}

\newcommand{\al}{\alpha}

\newcommand{\la}{\lambda}

\newcommand{\s}{\sigma}

\newcommand{\lb}{\label}
\newcommand{\rf}{\ref}
\newcounter{cnt1}
\newcounter{cnt2}
\newcounter{cnt3}
\newcommand{\blr}{\begin{list}{$($\roman{cnt1}$)$}
 {\usecounter{cnt1} \setlength{\topsep}{0pt}
 \setlength{\itemsep}{0pt}}}
\newcommand{\bla}{\begin{list}{$($\alph{cnt2}$)$}
 {\usecounter{cnt2} \setlength{\topsep}{0pt}
 \setlength{\itemsep}{0pt}}}
\newcommand{\bln}{\begin{list}{$($\arabic{cnt3}$)$}
 {\usecounter{cnt3} \setlength{\topsep}{0pt}
 \setlength{\itemsep}{0pt}}}
\newcommand{\el}{\end{list}}
\newtheorem{thm}{Theorem}[section]

\newtheorem{Def}[thm]{Definition}

\newtheorem{rem}[thm]{Remark}
\newcommand{\Rem}{\begin{rem} \rm}
\newcommand{\bdfn}{\begin{Def} \rm}
\newcommand{\edfn}{\end{Def}}

\newcommand{\ba}{\begin{array}}
\newcommand{\ea}{\end{array}}

\numberwithin{equation}{section}

\begin{document}

%\preprint{}

\title{Dual Relativistic Quantum Mechanics I}


\author{Tepper L. Gill}
\email[]{tgill@howard.edu}
\affiliation{ EECS,  Mathematics and Computational Physics Laboratory,\\ 
Howard University Washington DC 20059 USA}

\author{Gonzalo Ares de Parga}
\email[]{gadpau@hotmail.com}
\affiliation{Departmento de F\'{\i}sica, Escuela Superior de F\'{\i}sica y
Matem\'{a}ticas, Instituto Polit\'{e}cnico Nacional COFAA;
Edif 9, U. P. Adolfo L\'{\o}pez Mateos, Zacatenco, Lindavista, 07738,
M\'{e}xico D.F., M{\'e}xico}

\author{Trey  Morris}
\email[]{Morris.Trey.J@gmail.com}
\affiliation{Department of EECS, 
Howard University Washington DC 20059 USA}

\author{Mamadou  Wade}
\email[]{mamadou.wade@howard.edu}
\affiliation{Department of EECS, 
Howard University Washington DC 20059 USA}

\date{\today}

\begin{abstract}
It was shown in \cite{1} that the ultra-violet divergence in quantum electrodynamics (QED) is caused by a violation of the time-energy uncertainly relationship, due to the implicit assumption of infinitesimal time information (Dyson's conjecture). In \cite{2} it was shown that Einstein's special theory of relativity and Maxwell's field theory have mathematically equivalent dual versions.  The dual versions arise from an identity relating observer time to proper time as a contact transformation on configuration space, which leaves phase space  invariant.  The special theory has a dual version in the sense that, for any set of $n$ particles, every observer has two unique sets of global variables $({\bf{X}}, t)$ and $({\bf{X}}, \tau)$ to describe the dynamics, where ${\bf{X}}$ is the (unique) canonical center of mass. In the  $({\bf{X}}, t)$ variables, time is relative and the speed of light is unique, while in the $({\bf{X}}, \tau)$ variables, time is unique and the speed of light is relative with no upper bound.  In the Maxwell case, the two sets of particle wave equations are not equivalent. The dual version contains an additional longitudinal (dissipative) radiation term that appears instantaneously with acceleration, leading to the prediction that radiation from a betatron (of any frequency) will not produce photoelectrons.  A major outcome is the dual unification of Newtonian mechanics and classical electrodynamics with Einstein's special theory of relativity, without a self-energy divergency, or need of the problematic Lorentz-Dirac equation or any assumptions about the size or structure of a particle.  The purpose of this paper is to introduce and develop the dual theory of relativistic quantum mechanics. We obtain three distinct dual relativistic wave equations that reduce to the Schr{\"o}dinger equation when minimal coupling is turned off. We show that the dual Dirac equation provides a new formula for the anomalous magnetic moment of a charged particle.  We can obtain the exact value for the electron g-factor and phenomenological values for the muon and proton g-factors.
\end{abstract}

\maketitle
\section*{\bf{Introduction}}
In classical electrodynamics, Dirac partially by-passed many of the problems left over from the nineteenth century by replacing particles by fields (see \cite{3}). This approach led to the first example of a divergent theory (infinite self-energy). Dirac showed that, by using both advanced and retarded fields and a limiting procedure, one obtained a dissipative term, which accounted for the radiation reaction problem as an addition to the Lorentz equation (Lorentz-Dirac equation). This self-energy divergency was the main motivation for the Wheeler-Feynman approach to classical electrodynamics (see \cite{4}). Their theory gave the same dissipative term while avoiding the self-energy divergency.  This approach could not be quantized, but provided insight for Feynman's approach to QED. 

The failure to directly solve the classical problems forced researchers to use the Dirac theory as the basis for relativistic quantum mechanics and QED. This program maintained the self-energy divergence and introduced a few others. These divergences were later by-passed by Feynman, Schwinger and Tomonaga in the late 1940's leading to the great successes of that era. Neither Feynman, Schwinger or Tomonaga  considered their work a complete theory or final solution. Their methods did not account for the full spectra Hydrogen, but required the solution of the eigenvalue problem from the Dirac equation as initial input.  

The predominant belief at the time was that they were on the right track, and  the remaining difficulties would eventually be cleaned up by the mathematicians.  However, the major mathematical investigations were restricted to the limited task of  providing justification for the subtraction methods used to handle the divergencies.  This justification never came and by the early 1980's, it became clear that it never would.  The development of the electro-weak theory and the standard model each added additional problems, caused by extensions of QED methods to higher energy scales.

Another (less known) line of investigation sought to directly deal with the physical cause for these problems based on a number of suggestions from Dirac, Dyson and Feynman (see \cite{1}). A major outcome was that the ultra-violet divergency came from a violation of the time-energy uncertainty relationship (as suggested by Dyson) and was  not a hint of some (unknown) deeper problems as many believed. 

The success of this and other efforts suggested that an investigation into the physical justification for time as a forth coordinate was in order.  The lack of justification resulted in the discovery of the dual theory of special relativity \cite{2} and the dual Maxwell theory \cite{6,7}.  The dual Maxwell theory identifies radiation reaction as a dissipative term in the ${\bf{E}}$ field equation.  This led to the elimination of three major problems with the Dirac version of classical electrodynamics: the self-energy divergence disappeared, the need for point particles disappeared and, the need for the (problematic) Lorentz-Dirac equation disappeared.  In addition, it was shown that the dissipative term is equivalent to  a small (dynamical) mass for the photon.  This latter property implies that a quantum field based on the dual  Maxwell theory will not lead to an infrared-divergence (see \cite{6}).

The purpose of this paper is to introduce the dual relativistic quantum theory. After a brief review of the dual single particle and Maxwell theory in the second section, we introduce the dual  relativistic quantum theory in the third section.  The fourth section is devoted to the dual Dirac equation,  which provides a new formula for the g-factor of the electron and can also be used to obtain exact (phenomenological) values for the muon and proton g-factors. 
\section{\bf{Dual Classical Theory}}
\subsection{Particle Clock}
To develop the dual classical theory, we assume a classical interacting particle defined on phase space with variables $(t, {\bf{x}}, {\bf{p}})$ and Hamiltonian $H$ as seen by an observer $O$ in an inertial frame (in the standard setup).       If ${\mathbf{w}}$ is the particle velocity, let ${\gamma ^{ - 1}}\left( {\mathbf{w}}\right) = \sqrt {1 - {{{{\mathbf{w}}^2}} \mathord{\left/
 {\vphantom {{{{\mathbf{w}}^2}} {{c^2}}}} \right.
 \kern-\nulldelimiterspace} {{c^2}}}} $.  The classical proper time is defined by $d\tau  =\sqrt {1 - {{{{\mathbf{w}}^2}} \mathord{\left/
 {\vphantom {{{{\mathbf{w}}^2}} {{c^2}}}} \right.
 \kern-\nulldelimiterspace} {{c^2}}}}dt$,
\beqn\lb{1}
 {\mathbf{w}} = \frac{{d{\mathbf{x}}}}{{dt}},\quad d{\tau^2} = d{t^2} - \tfrac{1}{{{c^2}}}d{{\mathbf{x}}^2}   
\eeqn
Rearranging the last term, we get $ d{t^2} = d{\tau^2} + \tfrac{1}{{{c^2}}}d{{\mathbf{x}}^2}$, so
\begin{equation}\lb{2}
 cdt = \left( {\sqrt {{{\mathbf{u}}^2} + {c^2}} } \right)d\tau ,\quad {\mathbf{u}} = \frac{{d{\mathbf{x}}}}{{d\tau }}= {\gamma}({\bf{w}}){{\mathbf{w}}}.
\end{equation}
If we let $b={\sqrt {{{\mathbf{u}}^2} + {c^2}} }$, the first term in equation (\ref{2}) becomes $cdt =bd\tau$.  This leads to our first identity:
\begin{equation}\lb{3}
\frac{1}{c}\frac{d}{{dt}} \equiv \frac{1}{{{b}}}\frac{d}{{d{\tau}}}.
\end{equation}
This identity provides the correct way to define the relationship between the proper time and observer time for the particle.   If we apply the identity to ${\bf{x}}$, we obtain our second new identity, showing that the transformation leaves the configuration (or tangent) space invariant:
\begin{equation}\lb{4}
\frac{{{{\mathbf{w}}}}}{c} = \frac{1}{c}\frac{{d{{\mathbf{x}}}}}{{dt}} \equiv \frac{1}{{{b}}}\frac{{d{{\mathbf{x}}}}}{{d\tau }} = \frac{{{{\mathbf{u}}}}}{{{b}}}.
\end{equation}
The new particle coordinates are $({\bf{x}}, \tau)$.  In this representation, the position ${\bf{x}}$ is uniquely defined relative to $O$, while $\tau$ is uniquely defined by the particle.   The particle momentum can be represented as ${\bf{p}}=m \gamma({\bf{w}}){\bf{w}}=m{\bf{u}}$, where $m$ is the particle rest mass.  Thus, the phase space variables $({\bf{x}}, {\bf{p}})$, are left invariant.  For later use, we also have ${\gamma}({\bf{w}})=H/mc^2$.  This allows us to  also write  $d{\tau} = \left( {{{{m}{c^2}} \mathord{\left/ {\vphantom {{{m}{c^2}} {{H}}}} \right.  \kern-\nulldelimiterspace} {{H}}}} \right)dt$. 
\subsection{\bf Dual Particle Theory}
For compatibility with quantum theory, we require that any change of clocks  be canonical. The key concept to our approach is seen by examining the time evolution of a dynamical parameter $W({\bf{x}},{\bf{p}})$, via the standard formulation in terms of the Poisson brackets:
\beqn
\frac{{dW}}{{dt}} = \left\{ {H,W} \right\}.
\eeqn
To represent the dynamics via the proper time of the particle, we use the representation $d\tau  = ({{mc^2 } \mathord{\left/ {\vphantom {{mc^2 } H}} \right. \kern-\nulldelimiterspace} H})dt$,
so that:
\[
\frac{{dW}}{{d\tau }} = \frac{{dt}}{{d\tau }}\frac{{dW}}{{dt}} = \frac{H}
{{mc^2 }}\left\{ {H,W} \right\}.
\]
Using the invariant rest energy $mc^2$, we determine the canonical proper-time Hamiltonian $K$ such that:
\[
\left\{ {K,W} \right\} = \frac{H}{{mc^2 }}\left\{ {H,W} \right\},\quad \left. K \right|_{{\mathbf{p}} = 0}  = \left. H \right|_{{\mathbf{p}} = 0}  = mc^2. 
\]
From 
\[
\begin{gathered} 
  \left\{ {K,W} \right\} = \left[ {\frac{H}{{mc^2 }}\frac{{\partial H}}{{\partial {\mathbf{p}}}}} \right]\frac{{\partial W}}{{\partial {\mathbf{x}}}} - \left[ {\frac{H}
{{mc^2 }}\frac{{\partial H}}{{\partial {\mathbf{x}}}}} \right]\frac{{\partial W}}
{{\partial {\mathbf{p}}}} \hfill \\
  {\text{          }} = \frac{\partial }{{\partial {\mathbf{p}}}}\left[ {\frac{{H^2 }}
{{2mc^2 }} + a} \right]\frac{{\partial W}}{{\partial {\mathbf{x}}}} - \frac{\partial }
{{\partial {\mathbf{x}}}}\left[ {\frac{{H^2 }}{{2mc^2 }} + a'} \right]\frac{{\partial W}}
{{\partial {\mathbf{p}}}}, \hfill \\ 
\end{gathered} 
\]
we see that $a = a' = \tfrac{1}{2}mc^2$.  Thus, assuming no explicit time dependence, we have:
\beqn{\lb{6}}
K = \frac{{H^2 }}{{2mc^2 }} + \frac{{mc^2 }}{2},\quad {\text{and }}\quad \frac{{dW}}
{{d\tau }} = \left\{ {K,W} \right\}.
\eeqn 
Since $\tau$ remains invariant during interaction (minimal coupling), we assume  $K$ also remains invariant.  Thus, if $\sqrt {c^2 {\mathbf{p}}^2  + m^2 c^4 }  \to \sqrt {c^2 {\bs{\pi }}^2  + m^2 c^4 }  + V$, where $\bs{\pi}  = {\mathbf{p}} - \tfrac{e}{c}{\mathbf{A}}$, with $\bf A$ the vector potential and $V$ the potential energy.  In this case, $K$ becomes:
\[
K=\frac{{ {\bs{\pi }}^2}}
{{2m}} + mc^2  + \frac{{V^2 }}
{{2mc^2 }} + \frac{{V\sqrt {c^2  {\bs{\pi }}^2+ m^2 c^4 } }}
{{mc^2 }}.
\]
If we set $H_0=\sqrt {c^2 {\bs{\pi }}^2  + m^2 c^4 }$, use standard vector identities with ${\bs{\nabla}} \times \bs\pi=-\tfrac{e}{c}\bf{B}$, and compute Hamilton's equations, we get: 
\beqa
\frac{{d{\bf{x}}}}{{d\tau }} = \frac{{\partial K}}{{\partial {\bf{p}}}} = \frac{H}{{m{c^2}}}\left( {\frac{{{c^2}\bs{\pi} }}{{{H_0}}}} \right) = \frac{b}{c}\left( {\frac{{{c^2}\bs{\pi} }}{{{H_0}}}} \right) \Rightarrow \frac{{d{\bf{x}}}}{{d\tau }} = \frac{b}{c}\frac{{d{\bf{x}}}}{{dt}}
\eeqa
and
\begin{align}
  &\frac{{d{\mathbf{p}}}}{{d\tau }} = \frac{b}{c}\frac{{\left[ {\left( {{c^2}{\bs{\pi }}  \cdot {\bs{\nabla}} } \right){\mathbf{A}} + \tfrac{e}{b}\left( {{c^2}{\bs{\pi }}  \times {\mathbf{B}}} \right)} \right]}}{{{H_0}}} - \frac{b}{c}{\bs{\nabla}} V \nonumber \\
  & = \frac{b}{c}\left[ {\left( {{\mathbf{u}} \cdot {\bs{\nabla}} } \right){\mathbf{A}} + \tfrac{e}{b}\left( {{\mathbf{u}} \times {\mathbf{B}}} \right)} \right] - \frac{b}{c}{\bs{\nabla}} V\nonumber \\
  \quad\label{2.6}\\
  & = \frac{b}{c}\left[ {e{\mathbf{E}} + \tfrac{e}{b}\left( {{\mathbf{u}} \times {\mathbf{B}}} \right) + \tfrac{e}{b}\frac{{d{\mathbf{A}}}}{{d\tau }}} \right]\quad  \Rightarrow  \nonumber \\
&\frac{c}{b}\frac{{d{\bs\pi} }}{{d\tau }}
= \left[ {e{\mathbf{E}} + \tfrac{e}{b}\left( {{\mathbf{u}} \times {\mathbf{B}}} \right)} \right]=\left[ {e{\mathbf{E}} + \tfrac{e}{c}\left( {{\mathbf{w}} \times {\mathbf{B}}} \right)} \right]=\frac{{d{\bs\pi} }}{{dt }}. \nonumber
\end{align}
The above shows that the standard and dual equations of motion are mathematically equivalent. (They are clearly not physically equivalent.) 
\subsection{Dual Maxwell Theory}  
To study the field of a particle, we write Maxwell's equations (in c.g.s. units): 
\begin{align}
&\nabla  \cdot {\mathbf{B}} = 0,\quad\quad \quad \quad \nabla  \cdot {\mathbf{E}} = 4\pi \rho , \nonumber \\
&\quad \label{2.7}\\
&\nabla  \times {\mathbf{E}} =  - \frac{1}{c}\frac{{\partial {\mathbf{B}}}}{{\partial t}},\quad \nabla  \times {\mathbf{B}} = \frac{1}{c}\left[ {\frac{{\partial {\mathbf{E}}}}
{{\partial t}} + 4\pi \rho {\mathbf{w}}} \right]. \nonumber 
\end{align} 
Using equations (\ref{3}) and (\ref{4}), we have ({{\it{the mathematically identical  representation}}):
\begin{align}
&\nabla  \cdot {\mathbf{B}} = 0,\quad \quad \quad \nabla  \cdot {\mathbf{E}} = 4\pi \rho , \nonumber \\
&\quad \label{2.8}\\
&\nabla  \times {\mathbf{E}} =  - \frac{1}{b}\frac{{\partial {\mathbf{B}}}}{{\partial \tau }},\quad \nabla  \times {\mathbf{B}} = \frac{1}{b}\left[ {\frac{{\partial {\mathbf{E}}}}
{{\partial \tau }} + 4\pi \rho {\mathbf{u}}} \right]. \nonumber 
\end{align}
Thus, we obtain a mathematically equivalent set of Maxwell's equations using the local time of the particle to describe its fields.  

To derive the corresponding wave equations, we  take the curl of the last two equations in (\ref{2.8}), and use standard vector identities, to get: 
\begin{align}
& \frac{1}{{b^2 }}\frac{{\partial^2 {\mathbf{B}}}}{{\partial \tau ^2 }} - \frac{{{\mathbf{u}} \cdot {\mathbf{a}}}}{{b^4 }}\left[ {\frac{{\partial {\mathbf{B}}}}
{{\partial \tau }}} \right] - \nabla ^2 \cdot {\mathbf{B}} = \frac{1}{b}\left[ 4\pi \nabla  \times (\rho {\mathbf{u}}) \right], \nonumber \\
&\quad \label{2.10}\\
& \frac{1}{{b^2 }}\frac{{\partial ^2 {\mathbf{E}}}}{{\partial \tau ^2 }} - \frac{{{\mathbf{u}} \cdot {\mathbf{a}}}}{{b^4 }}\left[ {\frac{{\partial {\mathbf{E}}}}
{{\partial \tau }}} \right] - \nabla ^2  \cdot {\mathbf{E}}=\\
& - \nabla (4\pi \rho ) - \frac{1}{b}\frac{\partial }{{\partial \tau }}\left[ {\frac{{4\pi (\rho {\mathbf{u}})}}{b}} \right], \nonumber \end{align}
where ${\bf{a}} = d{\bf{u}}/d\tau$ is the particle acceleration.  The new term in equation (\ref{2.10}) is dissipative, acts to oppose the acceleration, is zero when ${\bf{a}} =0$  or perpendicular to $\bf{u}$ and  arises instantaneously with the force.  This makes it clear that the local clock encodes information about the particle's interaction that is unavailable when the clock of the observer or co-moving observer is used to describe the fields.   Furthermore,  this term does not depend on the nature of the force.  This is exactly what one expects of the back reaction caused by inertial resistance of the particle to accelerated motion and, according to Wheeler and Feynman \cite{4}, is precisely what is meant by radiation reaction. It follows that no consideration of the action of a particle on itself or the problematic Lorentz-Dirac equation is required to account for radiation reaction.

The $\bf{E}$ and $\bf{B}$ fields can be computed in the standard manner (using only retarded potentials) to get: (see \cite{5})
\beqa
\begin{gathered}
  {\mathbf{E}}\left( {{\mathbf{x}},\tau } \right) = \frac{{q{{\mathbf{r}}_u}\left( {1 - {{\mathbf{u}}^2}/{b^2}} \right)}}{{{s^3}}} + \frac{{q\left[ {{\mathbf{r}} \times \left( {{{\mathbf{r}}_u} \times {\mathbf{a}}} \right)} \right]}}{{{b^2}{s^3}}} \hfill \\
   + \frac{{q\left( {{\mathbf{u}} \cdot {\mathbf{a}}} \right)\left[ {{\mathbf{r}} \times \left( {{\mathbf{u}} \times {\mathbf{r}}} \right)} \right]}}{{{b^4}{s^3}}} \hfill \\ 
\end{gathered}
\eeqa
and
\beqa
\begin{gathered}
  {\mathbf{B}}\left( {{\mathbf{x}},\tau } \right) = \frac{{q\left( {{{\mathbf{r}}_u} \times {\mathbf{r}}} \right)\left( {1 - {{\mathbf{u}}^2}/{b^2}} \right)}}{{r{s^3}}} + \frac{{q{\mathbf{r}} \times \left[ {{\mathbf{r}} \times \left( {{{\mathbf{r}}_u} \times {\mathbf{a}}} \right)} \right]}}{{r{b^2}{s^3}}} \hfill \\
   + \frac{{qr\left( {{\mathbf{u}} \cdot {\mathbf{a}}} \right)\left( {{\mathbf{r}} \times {\mathbf{u}}} \right)}}{{{b^4}{s^3}}}. \hfill \\ 
\end{gathered}
\eeqa
It is easy to see that ${\bf B}$ is orthogonal to ${\bf E}$. The last term in each case arises because of the dissipative terms in the respective equation.  These  terms are zero if $\bf a$ is zero or orthogonal to $\bf u$.  In the first case, there is no radiation and the particle moves with constant velocity so that the field is massless. The second case depends on  the creation of motion which keeps $\bf a$ orthogonal to $\bf u$ (for example a betatron).   Since  ${\bf r}\times \left( {{\bf u}\times {\bf r}}
\right)=r^2{\bf u}-\left( {{\bf u}\cdot {\bf r}} \right){\bf r}$, we see that there is a
component along the direction of propagation (longitudinal).  (Thus,  the $\bf E$ field has a  longitudinal part.) This shows that the new dissipative term is equivalent to an effective mass,  meaning that the cause for radiation reaction comes directly from the use of the local clock to formulate Maxwell's equations.  Thus, there is no need to assume advanced potentials, self-interaction or mass renormalization along with the Lorentz-Dirac equation in order to account for radiation reaction as is done in Dirac's theory. Furthermore, no assumptions about the structure of the source are required.
\begin{rem} 
We conjecture that the above effective mass is the actual source of the photoelectric effect and that the photon is a real particle of non-zero (dynamical) mass, which travels with the fields.  If this conjecture is correct, radiation from a betatron (of any frequency)  exposed to a metal surface will not produce photo electrons.
\end{rem}  
\section{{Dual Relativistic Quantum Theory}}
The Klein-Gordon and Dirac equations were first discovered in early  attempts to make quantum mechanics compatible with the Minkowski formulation of the special theory of relativity.   Both were partially successful but could no longer be interpreted as particle equations.  A complete solution required quantum field theory and its associated problems.  In this section we introduce the dual relativistic quantum theory, which always has a single particle theory. 

Using equation (\ref{6}),  we follow the standard procedure to quantize leading to:
\[
i\hbar \frac{{\partial \Phi }}{{\partial \tau }} = K\Phi  = \left[ {\frac{{{H^2}}}{{2m{c^2}}} + \frac{{m{c^2}}}{2}} \right]\Phi .
\]
In addition to the Dirac Hamiltonian, there are two other possible Hamiltonians, depending on the way the potential appears with the square-root operator:
\[
{\bs{\beta}} \sqrt {{c^2}{\bs{\pi}}^2  - ec\hbar{\bs{\Sigma}}   \cdot {\mathbf{B}} + {m^2}{c^4}}  + V
\]
and
\[
{\bs{\beta}}{} \sqrt {{c^2}{\bs{\pi}}^2  - ec\hbar {\bs{\Sigma}}  \cdot {\mathbf{B}} + {{\left( {m{c^2} +{\bs{\beta}} V} \right)}^2}}. 
\] 
This gives us three possible dual relativistic particle equations for spin-$\tfrac{1}{2}$ particles (see also \cite{7}).   
\begin{enumerate}
\item The dual  Dirac equation:  
\beqn
\begin{gathered}
  i\hbar \frac{{\partial \Psi }}{{\partial \tau }} = \left\{ {\frac{{{\bs{\pi}^2}}}{{2m}} + \bs{\beta} {V} + m{c^2}} \right. - \frac{{e\hbar \bs{\Sigma}  \cdot {\mathbf{B}}}}{{2mc}} \hfill \\
  \quad \quad \;\;\left. { + \frac{{{V} \bs{\alpha}  \cdot \bs{\pi} }}{{mc}} - \frac{{i\hbar \bs{\alpha}  \cdot \nabla {V}}}{{2mc}} + \frac{{{V^2}}}{{2{mc^2}}}} \right\}\Psi . \hfill \\ 
\end{gathered} 
\eeqn
\item The dual version of the square-root equation, using the first possibility:
\beqn
\begin{gathered}
i\hbar \frac{{\partial \Psi }}{{\partial \tau}} = \left\{ \frac{{{\bs{\pi}}^2 }}{{2m}} - \frac{{e\hbar \bs{\Sigma}  \cdot {\bf{B}}}}{{2mc }} + mc^2  + \frac{{V^2 }}{{2mc^2 }}\right\}\Psi  \hfill \\
+ \frac{{V{\bs{\beta}} \sqrt {c^2 {\bm{\pi}}^2  - ec\hbar \bs{\Sigma}  \cdot {\bf{B}} + m^2 c^4 } }}{{2mc^2 }}\Psi \hfill \\
+ \frac{{{\bs{\beta}} \sqrt {c^2 {\bs{\pi}}^2  - ec\hbar \bs{\Sigma}  \cdot {\bf{B}} + m^2 c^4 } }}{{2mc^2 }}V \Psi. \hfill \\
\end{gathered} 
\eeqn
\item The dual version of the square-root equation, using the second possibility:
\beqn
\begin{gathered}
  i\hbar \frac{{\partial \Psi }}{{\partial \tau }}= \hfill \\
 \{\frac{{ {\bs{\pi}}^2 }}{{2m}} + {{\bs{\beta}}}V + m{c^2} - \frac{{e\hbar \bs{\Sigma}   \cdot {\mathbf{B}}}}{{2mc}} + \frac{{{V^2}}}{{2m{c^2}}}\}\Psi. \hfill \\
\end{gathered}   
\eeqn
\end{enumerate} 
If $\bf{A}$ and $V$ are zero, all equations reduce to:
\[
i\hbar \frac{{\partial \Psi }}{{\partial \tau }} = \left\{ {\frac{{{\bf{p}^2}}}{{2m}} + m{c^2}}  \right\}\Psi,
\]
which is the Schr\"{o}dinger equation with an added mass term.  This makes it easy to see that, in all cases, $K$ is positive definite.  In mathematical terms, the lower order terms are relatively bounded with respect to ${\bf{p}}^2/{2m}$.  It follows that, unlike the Dirac  and Klein-Gordon approach, we can interpret these equations as representations for actual particles. In the above equations, we have assumed that $V$ is time independent.  (However,  since ${\mathbf{A}}({\mathbf{x}},\tau)$ can have general time-dependence,  $\sqrt {c^2 {\bm{\pi}}^2  - ec\hbar {\bs{\Sigma}}  \cdot {\bf{B}} + m^2 c^4 }$ need not be related to the Dirac operator by a Foldy-Wouthuysen type transformation.)

\section{{The Dual Dirac Theory}}
We restrict our investigation to the dual Dirac equation.
Let ${\bf{s}}_p$ and ${\bs{\mu}} _p=2\mu_p{\bf{s}}_p$ be the proton spin and magnetic moment operators respectively. Let  $r_0=e^2/mc^2$ be the classical electron radius,  $\al  = \tf{{{e^2}}}{{\hbar c}}$  be the fine structure constant and let ${\bs{\alpha}}  = \left( {{\alpha _1},{\alpha _2},{\alpha _3}} \right) $ be the standard Dirac matrix, where ${\alpha _i} = \left[ {{\mathbf{0}},{\sigma _i},{\sigma _i},{\mathbf{0}}} \right]$,  
\[{\sigma _1} = \left( {\begin{array}{*{20}{c}}
  {\mathbf{0}}&1 \\ 
  1&{\mathbf{0}} 
\end{array}} \right),{\text{ }}{\sigma _2} = \left( {\begin{array}{*{20}{c}}
  {\mathbf{0}}&{ - i} \\ 
  i&{\mathbf{0}} 
\end{array}} \right),{\text{ }}{\sigma _3} = \left( {\begin{array}{*{20}{c}}
  {\mathbf{1}}&{\mathbf{0}} \\ 
  {\mathbf{0}}&{ - {\mathbf{1}}} 
\end{array}} \right)\]
and $\sigma=[\sigma_1, \sigma_2, \sigma_3]^t$.
The  potentials can be written as $ {V_0} = {{ - m{c^2}{r_0}} \mathord{\left/
 {\vphantom {{ - m{c^2}{r_0}} r}} \right.
 \kern-\nulldelimiterspace} r}  ,\quad      
{\bf{A}} = {{{{\bs{\mu}} _p}  \times {\bf{r}}} \mathord{\left/
 {\vphantom {{{{\bs{\mu}} _p}  \times {\bf{r}}} {{r^3}}}} \right.
 \kern-\nulldelimiterspace} {{r^3}}} $, where the spin orientation is along the z-axis (i.e., $A_r=A_\theta=0$ and $A_\phi=\tf{2\mu_p s_p \sin \theta}{r^2}$).  In what follows, ${\bs{\pi}} = \bf{p} - \tf{e}{c} \bf{A}$ and $\pi $ is the area of the unit circle.
\subsection{The Dirac Equation}
The eigenvalue problem for the Dirac equation $\la\Psi=H_D\Psi$, with $\Psi=[\psi_1,\psi_2]$, can be written as:
\beqn\lb{l1}
\begin{gathered} 
(\la - V - m{c^2}){\psi _1} = c(\sigma  \cdot \bs{\pi} ){\psi _2} \hfill \\
(\la - V + m{c^2}){\psi _2} = c(\sigma  \cdot \bs{\pi} ){\psi _1}. \hfill \\
\end{gathered}
\eeqn
Solving the second equation for  ${\psi _2}$  we have:
\beqn\lb{da}
{\psi _2} = c{\left[ {\lambda  - {V_0} + m{c^2}} \right]^{ - 1}}\left( {\sigma  \cdot {\bs{\pi }} } \right){\psi _1}
\eeqn
\subsection{The Dual Dirac Equation}
With $H_D = c\bs{\alpha}  \cdot {\bs{\pi }} + m{c^2}\beta + {V_0} =H_0 +V_0$, let $V = \tfrac{1}{{2m{c^2}}}\left[ {{H_0}{V_0} + {V_0}{H_0}} \right] $.  Then, we can write  the dual Dirac Hamiltonian as: 
\beqn
\begin{gathered} 
  K_D = \frac{{{H_D^2}}}{{2m{c^2}}} + \frac{{m{c^2}}}{2} = \frac{{{{\bs{\pi }}^2}}}{{2m}}+V - \frac{{e\hbar \Sigma  \cdot {\mathbf{B}}}}{{2mc}} + m{c^2} \hfill \\
   + \frac{{V_0^2}}{{2m{c^2}}}, \hfill \\
\end{gathered}   
\eeqn{\lb{dd}}  
\subsection{The Eigenvalue Problem} 
The general eigenvalue problem is:
\beqn{\lb{de}}
\begin{gathered}
  {{E \Psi }} = \left\{ {\frac{{{{\bs{\pi}}^2}}}{{2m}} + \beta {V_0} + m{c^2}} \right. - \frac{{e\hbar \Sigma  \cdot {\mathbf{B}}}}{{2mc}} \hfill \\
  \quad \quad \;\;\left. { + \frac{{{V_0}\alpha  \cdot \bs{\pi} }}{{mc}} - \frac{{i\hbar \alpha  \cdot \nabla {V_0}}}{{2mc}} + \frac{{{V_0^2}}}{{2{mc^2}}}} \right\}\Psi . \hfill \\ 
\end{gathered} 
\eeqn
Where, as before $\Psi=[\psi_1, \psi_2]^t$, with $\psi_1, \; \psi_2$ the upper and lower spinor components.  With  $\bf{A}=0$, and the exact eigenvalues for $\la \Psi=H_D\Psi$, we can use $\left[ {{\textstyle{{{\lambda ^2}} \over {2m{c^2}}}} + {\textstyle{{m{c^2}} \over 2}}} \right]\Psi  = {K_D}\Psi $ to find the exact eigenvalues for:
\[
\begin{gathered}
E\Psi  = \left\{ {\frac{{{{\mathbf{p}}^2}}}{{2m}} + \beta {V_0} + m{c^2} + \frac{{V_0^2}}{{2m{c^2}}} } \right\}\Psi \hfill \\
+ \left\{ \frac{{{V_0}\alpha  \cdot {\mathbf{p}}}}{{2m}} - \frac{{i\hbar \alpha  \cdot \nabla {V_0}}}{{2mc}}+{\frac{{{V_0}\alpha  \cdot {\mathbf{p}}}}{{2m}} - \frac{{i\hbar \alpha  \cdot \nabla {V_0}}}{{2mc}}} \right\}\Psi. \hfill \\
\end{gathered} 
\]
For further analysis, it is convenient to split (\rf{de}) into two equations:
\beqn\lb{d2}
\begin{gathered}
E\psi _1= \left\{ {\frac{{{{\bs{\pi}}^2}}}{{2m}} + V + m{c^2}- \frac{{e\hbar{\bs{\sigma}}  \cdot {\mathbf{B}}}}{{2mc}} + \frac{{{V_0^2}}}{{2m{c^2}}}} \right\}{\psi _1} \hfill \\
  {\text{             }} + \left\{ {\frac{{V_0{\bs{\sigma}}  \cdot \bs{\pi} }}{{mc}} - \frac{{i\hbar {\bs{\sigma}}  \cdot \nabla V_0}}{{2mc}}} \right\}{\psi _2} \hfill \\
E \psi _2= \left\{ {\frac{{{{\bs{\pi}}^2}}}{{2m}} - V + m{c^2} - \frac{{e\hbar {\bs{\sigma}}  \cdot {\mathbf{B}}}}{{2mc}} + \frac{{{V_0^2}}}{{2m{c^2}}}} \right\}{\psi _2} \hfill \\
  {\text{             }} + \left\{ {\frac{{V_0{\bs{\sigma}}  \cdot \bs{\pi} }}{{mc}} - \frac{{i\hbar {\bs{\sigma}}  \cdot \nabla V_0}}{{2mc}}} \right\}{\psi _1}. \hfill \\ 
\end{gathered}
\eeqn

 If we now use equation ({\rf{da}}), with the denominator to the left, we get: 
\[
{\psi _2} = \frac{{c{\bs{\sigma}}  \cdot {\bs{\pi}} }}{{\lambda  - {V_0} + m{c^2}}}{\psi _1}.
\]
We can now drop the second equation in (\rf{d2}) and convert the first to the stationary case and, (using $\psi$)   to get the eigenvalue equation:
\[\begin{gathered}
  E\psi  = \left\{ {\frac{{{{\bs{\pi}} ^2}}}{{2m}} + {V_0} + m{c^2} - \frac{{e\hbar {\bs{\sigma}}  \cdot {\mathbf{B}}}}{{2mc}} + \frac{{V_0^2}}{{2m{c^2}}}} \right\}\psi  \hfill \\
   + \left\{ {\frac{{{V_0}{\bs{\sigma}}  \cdot {\bs{\pi}} }}{{mc}} - \frac{{i\hbar {\bs{\sigma}}  \cdot \nabla {V_0}}}{{2mc}}} \right\}\frac{{c{\bs{\sigma}}  \cdot {\bs{\pi}} }}{{\left( {\lambda  - {V_0} + m{c^2}} \right)}}\psi.  \hfill \\ 
\end{gathered} \]
Expanding, we have: 
\beqn\lb{d4}
\begin{gathered}
  E\psi  =   \left\{ {\frac{{{{\bs{\pi}} ^2}}}{{2m}} + {V_0} + m{c^2} - \frac{{e\hbar {\bs{\sigma}}  \cdot {\mathbf{B}}}}{{2mc}} + \frac{{V_0^2}}{{2m{c^2}}}} \right\}\psi \hfill \\
   - \frac{{i\hbar \left( {{\bs{\sigma}}  \cdot \nabla {V_0}} \right)\left( {{\bs{\sigma}}  \cdot {\bs{\pi}} } \right)}}{{2m\left( {\lambda  - {V_0} + m{c^2}} \right)}}\psi  \hfill \\
   + \frac{{{V_0}}}{m}\frac{{\left( {{\bs{\sigma}}  \cdot {\mathbf{p}}{V_0}} \right)\left( {{\bs{\sigma}}  \cdot {\bs{\pi}} } \right)}}{{{{\left( {\lambda  - {V_0} + m{c^2}} \right)}^2}}}\psi  + \frac{{{V_0}}}{m}\frac{{\left( {{\bs{\sigma}}  \cdot {\bs{\pi}} } \right)\left( {{\bs{\sigma}}  \cdot {\bs{\pi}} } \right)}}{{\left( {\lambda  - {V_0} + m{c^2}} \right)}}\psi . \hfill \\ 
\end{gathered} 
\eeqn
Since the binding energy in Hydrogen is $13$ev and the rest mass of the electron is $5 \times 10^{5}$ev, the ratio is $2.6 \times 10^{-5}$.  Thus, there is little loss if we replace $\la - V + m{c^2}$ by $2m{c^2} + \tfrac{{{e^2}}}{r}$ in equation (\rf{da}). This allows us to by-pass the non-linear eigenvalue problem but we must still impose a cut-off since since the denominator is undefined at $r=0$.  With $r_0=\tfrac{{{e^2}}}{m{c^2}}$, we can write (\rf{da}) as:
\beqn{\lb{l3}}
{\psi _2} = \frac{{c(\bs{\sigma}  \cdot \bs{\pi} )}}{{2m{c^2}\left( {1 + \tfrac{{{r_0}}}{{2r}}} \right)}}{\psi _1}.
\eeqn
Using this, equation (\rf{d4}) becomes:
\beqn\lb{d5}
\begin{gathered}
  E\psi  = \left\{ {\frac{{{{\bs{\pi}} ^2}}}{{2m}} + {V_0} + m{c^2} - \frac{{e\hbar {\bs{\sigma}}  \cdot {\mathbf{B}}}}{{2mc}} + \frac{{V_0^2}}{{m{c^2}}}} \right\}\psi \hfill \\
   - \frac{{i\hbar \left( {{\bs{\sigma}}  \cdot \nabla {V_0}} \right)\left( {{\bs{\sigma}}  \cdot {\bs{\pi}} } \right)}}{{4{m^2}{c^2}\left( {1 + \tfrac{{{r_0}}}{{2r}}} \right)}}\psi  \hfill \\
   + \frac{{{V_0}\left( {{\bs{\sigma}}  \cdot {\mathbf{p}}{V_0}} \right)\left( {{\bs{\sigma}}  \cdot {\bs{\pi}} } \right)}}{{{4m^3}{c^4}{{\left( {1 + \tfrac{{{r_0}}}{{2r}}} \right)}^2}}}\psi  + \frac{{{V_0}\left( {{\bs{\sigma}}  \cdot {\bs{\pi}} } \right)\left( {{\bs{\sigma}}  \cdot {\bs{\pi}} } \right)}}{{{2m^2}{c^2}\left( {1 + \tfrac{{{r_0}}}{{2r}}} \right)}}\psi . \hfill \\ 
\end{gathered} 
\eeqn
The terms inside the first brace are essentially the leading terms for the Schr{\"o}dinger equation (when $\bf{A}=0$).  For proof of concept, we will treat the remaining terms as a first order perturbation.
\subsubsection{The S-state Problem}
Our main interest is in the s-state spectra, but before proceeding, we need to calculate the terms which contain $\left( \bs{\sigma }\cdot 
\bm{\pi }\right) \left( \bs{\sigma }\cdot \bs{\pi }\right) $ and 
$\mathbf{-}i\hbar \left( \bs{\sigma }\cdot \bs{\nabla }V_{0}\right)
\left( \bs{\sigma \cdot \bs{ \pi} }\right) $. For this, we use
the relations%
\begin{equation}
\left( \bs{\sigma }\cdot \mathbf{X}\right) \left( \bs{\sigma }\cdot 
\mathbf{Y}\right) =\mathbf{X\cdot Y}+i\bs{\sigma \cdot }\left( \mathbf{%
X\times Y}\right) .  \label{47}
\end{equation}%Using ${\mathbf{X}}={\mathbf{Y}}={\bs{\pi}}$  in (\rf{d4}), we have:
If  $\mathbf{X=Y}=\bm{\pi }$, we have%
\begin{eqnarray*}
\left( \bs{\sigma }\cdot \bs{\pi }\right) \left( \bs{\sigma }%
\cdot \bs{\pi }\right)  &=&\bs{\pi }^{2}+i\bs{\sigma \cdot }%
\left( \bs{\pi \times \pi }\right)  \\
\bs{\pi \times \pi } &\mathbf{=}&\frac{ie\hbar }{c}\mathbf{B}
\end{eqnarray*}
\begin{equation}
\left( \bs{\sigma }\cdot \bs{\pi }\right) \left( \bs{\sigma }%
\cdot \bs{\pi }\right) =\bs{\pi }^{2}-\frac{e\hbar }{c}\bs{%
\sigma \cdot B.}  \label{48}
\end{equation}%
If $\mathbf{X=-}i\hbar \bs{\nabla }V_{0}$ and $\mathbf{Y}=\bs{\pi }$,
we have:
\begin{equation}
\left( \mathbf{-}i\hbar \bs{\sigma }\cdot \bs{\nabla }V_{0}\right)
\left( \bs{\sigma }\cdot \bs{\pi }\right) =\mathbf{-}i\hbar \bm{%
\nabla }V_{0}\cdot \bs{\pi +}i\bs{\sigma \cdot }\left( \mathbf{-}%
i\hbar \bs{\nabla }V_{0}\bm{\times \pi }\right) .  \label{49}
\end{equation}%
By using $\bs{\pi}=\left( \bf{p}-\frac{\mathbf{e}}{c}\mathbf{A}%
\right) $, we arrive at%
\[
\begin{gathered}
  \left( { - i\hbar \bs{\sigma }  \cdot \bs{\sigma } {V_0}} \right)\left( {\bs{\sigma }  \cdot \bs{\pi} } \right) \hfill \\
   =  - i\hbar \bs{\sigma } {V_0} \cdot \bs{\pi}  + \hbar \bs{\sigma }  \cdot \left( {\bs{\sigma } {V_0} \times \bs{\pi} } \right) \hfill \\
   =  - i\hbar \bs{\sigma } {V_0} \cdot {\mathbf{p}} + \hbar \bs{\sigma }  \cdot \left( {\bs{\sigma } {V_0}} \right) \times {\mathbf{p}} \hfill \\
   + \tfrac{{ie\hbar }}{c}\left[ {\left( {\bs{\sigma } {V_0} \cdot {\mathbf{A}}} \right) + i\bs{\sigma }  \cdot \left( {\bs{\sigma } {V_0} \times {\mathbf{A}}} \right)} \right]. \hfill \\ 
\end{gathered} 
\]
Since $\mathbf{A\propto e}_{\varphi }$, we see that $\bs{\nabla }V_{0}\cdot \mathbf{A\propto \mathbf{e}}_{r}\mathbf{\cdot e}%
_{\varphi }=0,\label{50a}$
so that
\beqn
\begin{gathered}
\left( \mathbf{-}i\hbar \bs{\sigma }\cdot \bs{\nabla }V_{0}\right)
\left( \bs{\sigma }\cdot \bs{\pi }\right) =\hfill \\ 
\mathbf{-}i\hbar \left( 
\bs{\nabla }V_{0}\cdot \mathbf{p}\right) +\bs{\hbar \sigma \cdot }%
\left( \bs{\nabla }V_{0}\mathbf{\times p}\right) \mathbf{-}\frac{\mathbf{%
e\hbar }}{c}\bs{\sigma \cdot }\left( \bs{\nabla }V_{0}\mathbf{\times
A}\right) \hfill \\ 
\end{gathered}
\eeqn
If we write $\mathbf{p}$ in spherical polar coordinates, we have:
\begin{equation}
\mathbf{p}=-i\hbar \bs{\nabla} =-i\hbar \left( \mathbf{e}_{r}\frac{%
\partial }{\partial r}+\frac{1}{r}\mathbf{e}_{\varphi }\frac{\partial }{%
\partial \varphi }+\frac{1}{\sin \theta }\mathbf{e}_{\theta }\frac{\partial 
}{\partial \theta }\right).   \label{51}
\end{equation}%
It then follows that:%
\begin{equation}
\mathbf{\nabla }V_{0}=\left( \mathbf{e}_{r}\frac{\partial V_{0}}{\partial r}+%
\frac{1}{r}\mathbf{e}_{\theta }\frac{\partial V_{0}}{\partial \theta }+\frac{%
1}{r\sin \theta }\mathbf{e}_{\varphi }\frac{\partial V_{0}}{\partial \varphi 
}\right) .  \label{52}
\end{equation}%
Then since $V_{0}=-\frac{e^{2}}{r}$, we have%
\beqn
\begin{gathered}
- i\hbar \nabla {V_0} \cdot {\mathbf{p}} =  \hfill \\
- i\hbar \left\{ {\frac{{\partial {V_0}}}{{\partial r}}{{\mathbf{e}}_r} \cdot \left[ { - i\hbar \left( {{{\mathbf{e}}_r}\frac{\partial }{{\partial r}}} \right)} \right]} \right\} = \tfrac{{{e^2}{\hbar ^2}}}{{{r^2}}}\frac{\partial }{{\partial r}}. \hfill \\ 
\end{gathered} 
\eeqn
We also have%
\begin{equation}
\mathbf{\hbar \sigma \cdot }\left( \mathbf{\nabla }V_{0}\mathbf{\times p}%
\right) =-\mathbf{\hbar }\frac{e^{2}}{r^{3}}\mathbf{\sigma \cdot L}
\label{54a}
\end{equation}%
Finnaly, with $\mathbf{A=}\tfrac{2\mu _{p}\left\vert s_{p}\right\vert \sin \theta }{r^{2}}
\mathbf{e}_{\varphi }$, we have 
\begin{equation}
\mathbf{-}\frac{\mathbf{e\hbar }}{c}\mathbf{\sigma \cdot }\left( \mathbf{%
\nabla }V_{0}\mathbf{\times A}\right) =-\frac{\mathbf{e\hbar }}{c}\frac{e^{2}%
}{r^{4}}2\mu _{p}\left\vert s_{p}\right\vert \sin \theta \left( \mathbf{%
\sigma \cdot e}_{\theta }\right) .  \label{56}
\end{equation}%
From these results, we can write the last three terms in equation (III.8) as:
\[\begin{gathered}
  \left( {\mathbf{a}} \right)\quad  - \frac{{i\hbar \left( {\sigma  \cdot \nabla {V_0}} \right)\left( {\sigma  \cdot \pi } \right)}}{{2{m^2}{c^2}\left( {1 + \tfrac{{{r_0}}}{{2r}}} \right)}} =  - \frac{{{e^2}{\hbar ^2}}}{{2{m^2}{c^2}\left( {1 + \tfrac{{{r_0}}}{{2r}}} \right){r^2}}}\frac{\partial }{{\partial r}} \hfill \\
  - \frac{{{e^2}\hbar }}{{2{m^2}{c^2}\left( {1 + \tfrac{{{r_0}}}{{2r}}} \right){r^3}}}\sigma  \cdot {\mathbf{L}} 
  - \frac{{{e^3}\hbar {\mu _p}\left| {{{\mathbf{s}}_p}} \right|}}{{{m^2}{c^3}\left( {1 + \tfrac{{{r_0}}}{{2r}}} \right){r^4}}}\sin \theta \left( {\sigma  \cdot {{\mathbf{e}}_\theta }} \right) \hfill \\
  \left( {\mathbf{b}} \right)\quad  + \frac{{{V_0}\left( {\sigma  \cdot \pi } \right)\left( {\sigma  \cdot \pi } \right)}}{{{m^2}{c^2}\left( {1 + \tfrac{{{r_0}}}{{2r}}} \right)}} =  - \frac{{{e^2}{\hbar ^2}{{\mathbf{p}}^2}}}{{{m^2}{c^2}r\left( {1 + \tfrac{{{r_0}}}{{2r}}} \right)}} \hfill \\
  - \frac{{{e^3}{\hbar ^2}{{\mathbf{A}}^2}}}{{{m^2}{c^3}r\left( {1 + \tfrac{{{r_0}}}{{2r}}} \right)}} + \frac{{{e^3}\hbar \sigma  \cdot {\mathbf{B}}}}{{{m^2}{c^3}r\left( {1 + \tfrac{{{r_0}}}{{2r}}} \right)}} \hfill \\
  \left( {\mathbf{c}} \right)\quad  + \frac{{{V_0}\left( {\sigma  \cdot {\mathbf{p}}{V_0}} \right)\left( {\sigma  \cdot \pi } \right)}}{{{m^3}{c^4}{{\left( {1 + \tfrac{{{r_0}}}{{2r}}} \right)}^2}}} =  - \frac{{{e^4}{\hbar ^2}}}{{{m^3}{c^4}{{\left( {1 + \tfrac{{{r_0}}}{{2r}}} \right)}^2}{r^3}}}\frac{\partial }{{\partial r}}\hfill \\
   - \frac{{{e^4}\hbar }}{{{m^3}{c^4}{{\left( {1 + \tfrac{{{r_0}}}{{2r}}} \right)}^2}{r^4}}}\sigma  \cdot {\mathbf{L}} 
 - \frac{{2{e^5}\hbar {\mu _p}\left| {{{\mathbf{s}}_p}} \right|}}{{{m^3}{c^5}{{\left( {1 + \tfrac{{{r_0}}}{{2r}}} \right)}^2}{r^5}}}\sin \theta \left( {\sigma  \cdot {{\mathbf{e}}_\theta }} \right). \hfill \\ 
\end{gathered} \]
When $\bf{A}=0$, these terms become:
\[\begin{gathered}
  \left( {{\mathbf{a'}}} \right)\quad  - \frac{{i\hbar \left( {\sigma  \cdot \nabla {V_0}} \right)\left( {\sigma  \cdot {\mathbf{p}}} \right)}}{{2{m^2}{c^2}\left( {1 + \tfrac{{{r_0}}}{{2r}}} \right)}} =  - \frac{{{e^2}{\hbar ^2}}}{{2{m^2}{c^2}\left( {1 + \tfrac{{{r_0}}}{{2r}}} \right){r^2}}}\frac{\partial }{{\partial r}}\hfill \\
   - \frac{{{e^2}\hbar }}{{2{m^2}{c^2}\left( {1 + \tfrac{{{r_0}}}{{2r}}} \right){r^3}}}\sigma  \cdot {\mathbf{L}} \hfill \\
  \left( {{\mathbf{b'}}} \right)\quad  + \frac{{{V_0}\left( {\sigma  \cdot {\mathbf{p}}} \right)\left( {\sigma  \cdot {\mathbf{p}}} \right)}}{{{m^2}{c^2}\left( {1 + \tfrac{{{r_0}}}{{2r}}} \right)}} =  - \frac{{{e^2}{\hbar ^2}{{\mathbf{p}}^2}}}{{{m^2}{c^2}r\left( {1 + \tfrac{{{r_0}}}{{2r}}} \right)}} \hfill \\
  \left( {{\mathbf{c'}}} \right)\quad  + \frac{{{V_0}\left( {\sigma  \cdot {\mathbf{p}}{V_0}} \right)\left( {\sigma  \cdot {\mathbf{p}}} \right)}}{{{m^3}{c^4}{{\left( {1 + \tfrac{{{r_0}}}{{2r}}} \right)}^2}}} =  - \frac{{{e^4}{\hbar ^2}}}{{{m^3}{c^4}{{\left( {1 + \tfrac{{{r_0}}}{{2r}}} \right)}^2}{r^3}}}\frac{\partial }{{\partial r}}\hfill \\
   - \frac{{{e^4}\hbar }}{{{m^3}{c^4}{{\left( {1 + \tfrac{{{r_0}}}{{2r}}} \right)}^2}{r^4}}}\sigma  \cdot {\mathbf{L}}. \hfill \\ 
\end{gathered} \]
The new terms that arise, separating equation (III.8) from the Schr{\"o}dinger equation, and when $\bf{A}\ne 0$ are the two terms from inside first brace of equation (III.8):
\beqa
\frac{{2{e^2}\mu _p^2{{\left| {{{\mathbf{s}}_p}} \right|}^2}{{\sin }^2}\theta }}{{m{c^2}{r^4}}} - \frac{{e\hbar \sigma  \cdot {\mathbf{B}}}}{{2mc}},
\eeqa
and the following three terms from above:
\beqa
\begin{gathered}
   - \frac{{{e^3}\hbar {\mu _p}\left| {{{\mathbf{s}}_p}} \right|}}{{{m^2}{c^3}\left( {1 + \tfrac{{{r_0}}}{{2r}}} \right){r^4}}}\sin \theta \left( {\sigma  \cdot {{\mathbf{e}}_\theta }} \right) - \frac{{4{e^3}{\hbar ^2}\mu _p^2{{\left| {{{\mathbf{s}}_p}} \right|}^2}{{\sin }^2}\theta }}{{{m^2}{c^3}{r^5}\left( {1 + \tfrac{{{r_0}}}{{2r}}} \right)}} \hfill \\
   + \frac{{{e^3}\hbar \sigma  \cdot {\mathbf{B}}}}{{{m^2}{c^3}r\left( {1 + \tfrac{{{r_0}}}{{2r}}} \right)}} 
   - \frac{{2{e^5}\hbar {\mu _p}\left| {{{\mathbf{s}}_p}} \right|}}{{{m^3}{c^5}{{\left( {1 + \tfrac{{{r_0}}}{{2r}}} \right)}^2}{r^5}}}\sin \theta \left( {\sigma  \cdot {{\mathbf{e}}_\theta }} \right). \hfill \\ 
\end{gathered}
\eeqa
Grouping and rearranging the terms, we have:
\beqn\lb{ls-a}
\tfrac{{4e{r_0}\hbar \sigma  \cdot {\mathbf{B}}}}{{2mc\left( {2r + {r_0}} \right)}} - \tfrac{{e\hbar \sigma  \cdot {\mathbf{B}}}}{{2mc}} = -\left[ {1 - \tfrac{{4{r_0}}}{{\left( {2r + {r_0}} \right)}}} \right]\tfrac{{e\hbar \sigma  \cdot {\mathbf{B}}}}{{2mc}},
\eeqn
\beqn\lb{ls}
\begin{gathered}
  \frac{{2{r_0}\mu _p^2{{\left| {{{\mathbf{s}}_p}} \right|}^2}{{\sin }^2}\theta }}{{{r^4}}} - \frac{{4e{r_0}{\hbar ^2}\mu _p^2{{\left| {{{\mathbf{s}}_p}} \right|}^2}{{\sin }^2}\theta }}{{mc{r^5}\left( {1 + \tfrac{{{r_0}}}{{2r}}} \right)}} \hfill \\
   = 2{r_0}\mu _p^2{\left| {{{\mathbf{s}}_p}} \right|^2}\left[ {1 - \frac{{4e{\hbar ^2}}}{{mc\left( {2r + {r_0}} \right)}}} \right]\frac{{{{\sin }^2}\theta }}{{{r^4}}} \hfill \\ 
\end{gathered} 
\eeqn
and
\beqn\lb{ls-1}
\begin{gathered}
   - \lt[\tfrac{{e{r_0}\hbar {\mu _p}\left| {{{\mathbf{s}}_p}} \right|}}{{mc\left( {1 + \tfrac{{{r_0}}}{{2r}}} \right){r^4}}}  + \tfrac{{2er_0^2\hbar {\mu _p}\left| {{{\mathbf{s}}_p}} \right|}}{{mc{{\left( {1 + \tfrac{{{r_0}}}{{2r}}} \right)}^2}{r^5}}}\rt]\sin \theta \left( {\sigma  \cdot {{\mathbf{e}}_\theta }} \right) \hfill \\
   =  - \frac{{2e{r_0}\hbar {\mu _p}\left| {{{\mathbf{s}}_p}} \right|}}{{mc\left( {2r + {r_0}} \right)}}\left[ {1 + \frac{{4r_0^{}}}{{\left( {2r + {r_0}} \right)}}} \right]\frac{{\sin \theta }}{{{r^3}}}\left( {\sigma  \cdot {{\mathbf{e}}_\theta }} \right) \hfill \\ 
\end{gathered}
\eeqn

In the remainder of the paper, we focus on the implications of equation (\rf{ls-a}) and the anomalous magnetic moment.  The implications of (\rf{ls}) and (\rf{ls-1}) will be a part of another study.
\subsection{Anomalous Magnetic Moment}
In this section, we investigate the equation (\rf{ls-a}) under the assumption that the charged, spin-$1/2$ particle does not possess any internal structure (a Dirac particle).  In this case, the spin magnetic moment is given by:
\[
\bs{\mu } = g\frac{{e }}{{2mc}}{\mathbf{s}}=g\mu_B{\mathbf{s}},
\]
where ${\mathbf{s}}=\tf{\hbar \s}{2}$ is the intrinsic spin operator.  We can also write the above as
\beqn{\lb{ls-ab}}
 H_{a} = 2\left[ {1 - \frac{{4{r_0}}}{{\left( {2r + {r_0}} \right)}}} \right]{\mu _B}{\mathbf{s}} \cdot {\mathbf{B}}
\eeqn

Thus, we have that:
\beqn
g_r =  2\left[ {1 - \frac{{4{r_0}}}{{\left( {2r + {r_0}} \right)}}} \right]
\eeqn
 If we take the cutoff at $r=\tf{r_0}{2}$, then $g=-2$, while if we take the cut off at $g = \mathop {\lim }\limits_{r \to 0} {g_r}$, we obtain $g=-6$.
Taking $r_e=   0.499857150068631 \times r_0$, we obtain the correct experimental result:
\[
 g=-2.00231930436256.
\]
If we treat the muon and proton phenomenologically we can also obtain their $g$-factors:
\beqn
\begin{gathered}
  g_\mu ^a = 2\left[ {1 - \frac{{4r_0^\mu }}{{\left( {2{r_\mu } + r_0^\mu } \right)}}} \right] \hfill \\
\quad \hfill \\
  g_p^a = -2\left[ {1 - \frac{{4r_0^p}}{{\left( {2{r_p} + r_0^p} \right)}}} \right], \hfill \\ 
\end{gathered}
\eeqn
where $r_0^\mu=\tf{e^2}{m_\mu c^2}$ and $r_0^p=\tf{e^2}{m_p c^2}$.
\section{Discussion}
At the classical level we find that the standard and dual theories are mathematically equivalent.  At the quantum level, the dual Dirac equation is not mathematically equivalent to the Dirac equation.  The dual Dirac equation is strictly positive definite, so that there are no problems with using it as a particle equation.  However, we must now directly face the existence of antiparticles. 
\smallskip
 
In order to do this, let us first revisit our conceptual view of the real numbers and their representation. Recall that a field is a set $\A$ that has two binary operations $\oplus$ and $\odot$ that satisfies all our common experience with real numbers.  Formally:
 \begin{Def}   The real numbers is a triplet $(\R, +, \cdot)$, which is a field, with $0$ as the additive identity $(i.e., a+0=a \; {\rm{for\, all}}\; a \in \R)$ and $1$ as the multiplicative identity $(i.e., \; a\cdot 1=a \; {\rm{for\, all}}\; a \in \R)$.
 \end{Def}
This structure was designed by mathematicians without regard to its possible use in physics.   Santilli \cite{8} defined the isodual number field for use in physics and that is what we need.  

 \begin{Def} The isodual real numbers $({\hat{\R}}, +, *)$ is a field, with $0=\hat{0}$ as the additive identity (i.e., $\hat{a}+\hat{0}=\hat{a}$ for all $-a=\hat{a} \in {{\hat{\R}}}$) and $\hat{1}=-1$ as the multiplicative identity (i.e., $\hat a * \hat 1 = ( - a)( - 1)( - 1) = \hat a$ for all $\hat{a} \in {\hat{\R}}$).
 \end{Def}
We note that we can obtain the isodual of any physical quantity $\hat{A}$ from the  equation  $A+\hat{A}=0$.

In our theory, the evolution of a particle is formally defined on a Hilbert space $\mathcal{H}$ over the complex numbers ${\C}={\R} + i{\R}$, with Hamiltonian $K$ by the equation
\[
i\hbar \frac{{\partial \psi }}{{\partial \tau}} = K\psi .
\]
The conjugate equation is:
\[ 
- i\hbar \frac{{\partial {\psi ^*}}}{{\partial \tau}} = K{\psi ^*}.
\]
If we use $\mathcal{\hat{C}}$ as our number field, we can write the above equation as:
\[
\hat{i}*\hat{\hbar} * \frac{{\partial {\psi ^*}}}{{\partial \hat \tau}} = {\hat{K}}*{\psi ^*}
\]
This approach allows us to naturally view anti-particles as particles with their proper time reversed and their evolution defined on $\mathcal{H^*}$ over $\mathcal{\hat{C}}$. (This does not imply that the time of the observer is reversed.)
\begin{rem}
Santilli \cite{8} has shown that charge conjugation and isoduallity are equivalent for  the particle-antiparticle symmetry operation.
\end{rem}
\section*{Conclusion}
In this paper we have introduced the dual relativistic quantum theory corresponding to Einstein's special theory of relativity and Maxwell's field theory \cite{2}.  The dual classical theory was shown to be mathematically equivalent,  but the dual quantum theory is not.  We have found three distinct dual relativistic wave equations that reduce to the Schr{\"o}dinger equation when minimal coupling is turned off.  We have focused on the dual Dirac equation and used it to derive a new formula for the g-factor of a spin-1/2 particle.  This allowed us to obtain the exact value for the electron g-factor.  The formula can also be applied to the muon and the proton.  Using the isodual numbers of Santilli \cite{8}, we have shown that our theory naturally interprets antiparticles as particles moving backwards in their proper time (and not the time of the observer).

\begin{acknowledgments}
We would like to thank Professors Netsivi Ben-Amots, Alexander Gersten, Larry Horwitz, Martin C. Land, Elliot Leib and Ruggero M. Santilli  for their continued interest, support and suggestions.  
\end{acknowledgments}

\bibliographystyle{apsrev4-2}

\end{document}